\documentclass[aps,11pt,showkeys,nofootinbib]{revtex4-2}

\usepackage{amsmath,amssymb,amsfonts,amsthm}
\usepackage{amsbsy} 
\usepackage{epsfig}
\usepackage{latexsym}
\input amssym.def
\input amssym.tex

\usepackage{hyperref}

\newcommand{\oarX}[1]{\href{http://arxiv.org/abs/#1}{{\ttfamily #1}}}
\newcommand{\arX}[1]{\href{http://arxiv.org/abs/#1}{{\ttfamily arXiv:#1}}}

\def\ben{\begin{equation}}
\def\een{\end{equation}}

\begin{document}

\title{Hilbert space formalisms for group field theory}

\author{Steffen Gielen}
\affiliation{School of Mathematical and Physical Sciences, University of Sheffield, Hicks Building, Hounsfield Road, Sheffield S3 7RH, UK}
\email{s.c.gielen@sheffield.ac.uk}

\date{\today}

%%%%%%%%%%%%%%%%%%%%%%%%%%%%%%%%%%%%%%%%%%%%%%%%%%%%%%%

\begin{abstract}
Group field theory is a background-independent approach to quantum gravity whose starting point is the definition of a quantum field theory on an auxiliary group manifold ({\em not} interpreted as spacetime, but rather as the finite-dimensional configuration space of a single ``atom'' of geometry). Group field theory models can be seen as an extension of matrix and tensor models by additional data, and are traditionally defined through a functional integral whose perturbative expansion generates a sum over discrete geometries. More recently, some efforts have been directed towards formulations of group field theory based on a Hilbert space and operators, in particular in applications to cosmology. This is an attempt to review some of these formulations and their main ideas, to disentangle these constructions as much as possible from applications and phenomenology, and to put them into a wider context of quantum gravity research.
\end{abstract}


\maketitle

\section{Introduction}

A general, perhaps surprising feature of quantum theories that describe Nature is that they can often be obtained via a process of ``quantisation'', first writing down a classical theory and then applying a set of rules to construct the corresponding quantum theory. This is the case for the theories underlying the Standard Model of particle physics, but also for theories beyond it such as inflation, or some effective theories used in condensed matter physics. The process of quantisation is usually a mixture of using well-established recipes and making specific choices, often informed by physical requirements, along the way. Traditionally quantisation proceeds either by canonical quantisation or by covariant (path integral) quantisation; more mathematically sophisticated methods such as geometric quantisation \cite{GeomQ} or deformation quantisation \cite{defQ} have also been developed. Usually one starts from a classical phase space, which defines classical observables and dynamics, and constructs the corresponding quantum observables, quantum dynamical equations, and so on. The path integral in its most general form is also an integral over phase space variables, although in the most common case, a Hamiltonian quadratic in momenta, one can integrate out momentum variables and reduce this to an integral over configuration space variables only \cite{Kleinert}. 
However, the path integral is often also defined directly in configuration space, as an integral over the set of dynamical configurations weighted by the exponential of a given action. This latter definition can be applied to systems which have no obvious phase space description, and hence appears more flexible. Approaches to quantum gravity such as causal dynamical triangulations \cite{CDT} and causal sets \cite{causalsetQ} use such ``pure Lagrangian'' definitions of discrete path integrals, with no phase space description used. For spin foam models (candidates for a quantisation of gravity or topological BF theory), one can start with a phase space definition which, regularised on a triangulation, results in a discrete sum over amplitudes \cite{SF}. 

Here we will focus on {\em group field theory} (GFT), often seen as an extension or completion of the spin foam approach to quantum gravity \cite{GFT10q} in the following sense: a group field theory is an interacting field theory defined on some group manifold, whose perturbative expansion generates a sum over spin foam amplitudes; it hence defines a certain spin foam model, or proposal for a discrete quantum gravity path integral. \cite{GFTfreidel} gives an introduction into the main ideas and a discussion of other properties of spin foam models that can be given a definition via group field theory, such as the physical inner product between boundary states, observables associated to boundary spin networks, and interpretation of ``bubble'' divergences in Feynman graphs. There are many reviews on group field theory (see also, e.g., \cite{GFTreviews}) and we will illustrate the main ideas in the simplest and perhaps best-known example of the Boulatov model below, but we want to focus on a more specific question: {\em given} a group field theory model constructed via spin foam methods or by other means, can we define its canonical quantisation (or, more generally speaking, a Hilbert space formalism)? This is a tricky issue since, even though group field theories are defined in terms of a standard Lagrangian, the obvious strategy of passing to a Hamiltonian via Legendre transform does not work due to the absence of a background time parameter -- recall that such a theory does not live on spacetime but on an abstract group manifold. Indeed, for many years the literature was exclusively formulated in path integral language and with connection to spin foam models, borrowing interpretational ideas such as the identification of boundary data with loop quantum gravity spin networks but not deriving these in a systematic way.

The first proposal for a Hilbert space formalism was laid out in \cite{Oriti2q}, applying the idea of second quantisation as used in condensed matter physics to spin networks of loop quantum gravity and showing how a certain kinematical Hilbert space could be constructed and given an interpretation in terms of (canonical) loop quantum gravity states. There was also a proposal for implementing dynamics, in the form of a constraint or projector on such spin-network states via a generalised partition function. This formalism was immediately applied to cosmology in \cite{GFTcondensate} focussing on coherent states and a mean-field approximation, proposing to use Schwinger--Dyson equations to extract dynamical information in an approximate sense. These ideas led to a new research programme connecting group field theory to cosmology \cite{GFTcondensatereview}, with follow-up papers proposing new ideas for extracting cosmological dynamics from such a Hilbert space formalism; a key step in this development was the extension of group field theory to models including a scalar matter field coupled to quantum gravity, which allowed for the definition of a type of relational dynamics and effective Friedmann equations \cite{GFTcosmorel}. Direct comparison of effective equations with other approaches such as loop quantum cosmology was now possible. At the same time, there were discussions of technical issues such as divergences and ideas for how to remedy those (see, e.g., \cite{thermalGFT}), but without clear connection to other examples of canonical quantisation and again mostly with cosmological applications and semiclassical approximations in mind. As a parallel development, a more conventional canonical quantisation was proposed  using the scalar matter field as a time coordinate \cite{relHamilt}, a very different approach from the one of \cite{Oriti2q}.

The aim of this review is to take a step back and give an overview over some of the ideas for Hilbert space formalisms for group field theory that have appeared in the recent literature, and to add some new technical comments and ideas that have not been discussed before. Much of the literature focuses on extracting phenomenology and presents particular state choices and approximation methods required to obtain effective semiclassical dynamics, which tends to come at the expense of a basic introduction to the theory. Technical discussions of important issues can be found in the literature but are spread over partial discussions in various papers (or even unpublished work). New models and applications are being proposed regularly, making it even more difficult to keep track of all of the separate ideas and their connection. We will refrain from discussing particular models in much detail or asking about the evidence supporting these as possible theories of quantum gravity; we will mention applications to cosmology only in a very minimal way; and we will not discuss results obtained in a covariant setting (such as the many results obtained in understanding the phase structure \cite{LandauGinz} and renormalisation \cite{renormreview} of these theories) other than to motivate specific models. We hope that this review will be useful for researchers in quantum gravity or quantum field theory who are trying to get an idea of current research in the rapidly evolving field of group field theory. With its more specific focus compared to other reviews, it might also serve as a resource for young researchers beginning to work in this exciting research area. Finally, we hope that it will also help to support dialogue within the community.

\section{What actually is a GFT?}
Let us finally introduce the definition of a group field theory. We first illustrate it with the example of the Boulatov model \cite{boulatov}, in its most interesting definition for discrete quantum gravity. The model is defined in terms of the {\em group field} $\phi$, which is a real-valued scalar field on ${\rm SU}(2)^3$:
\ben
\phi: {\rm SU}(2)^3 \rightarrow \mathbb{R}\,,\quad \phi(g_1,g_2,g_3)=\phi(g_1h,g_2h,g_3h)\quad\forall h\in {\rm SU}(2)\,,\een
\ben\phi(g_1,g_2,g_3)=\phi(g_2,g_3,g_1)=\phi(g_3,g_1,g_2)\,,
\een
\ben
\label{boulatovaction}
S[\phi]=\frac{1}{2}\int{\rm d}^3 g\;\phi^2(g_1,g_2,g_3)-\frac{\lambda}{4!}\int {\rm d}^6 g \;\phi(g_1,g_2,g_3)\phi(g_1,g_4,g_5)\phi(g_2,g_5,g_6)\phi(g_3,g_6,g_4)\,.
\een
The group field is required to have a right-translation symmetry and a permutation symmetry; the action $S[\phi]$ consists of a quadratic (``kinetic'') term and a quartic, ``combinatorially nonlocal'' interaction. Notice that the combinatorial pattern in the interaction term is precisely that needed to glue four triangles along six shared edges to form a tetrahedron. The model can be generalised by replacing ${\rm SU}(2)$ by a general compact group $G$, as we will do later on.

With those definitions Boulatov showed that
\ben
\int \mathcal{D}\phi\;e^{-S[\phi]} = \sum_C \lambda^{N_3(C)}\left( \sum_{\{j_e\}}\prod_{e\in C}(2j_e+1) \prod_{T\in C}\{6j\}_T \right)
\label{GFTpert}
\een
where $C$ is an oriented three-dimensional simplicial complex, $N_3(C)$ is the number of tetrahedra in $C$, $\{j_e\}$ stands for a possible assignment of irreducible ${\rm SU}(2)$ representations to all edges $e\in C$, $T\in C$ is a tetrahedron in $C$ and $\{6j\}_T$ stands for the $6j$-symbol with entries the $j_e$ assigned to the six edges of each $T$. For given $C$, the expression in brackets corresponds to the amplitude assigned to $C$ in the {\em Ponzano--Regge model} for quantum gravity in three dimensions \cite{PonzanoRegge} (see \cite{Barrett} for many technical subtleties regarding the Ponzano--Regge model, which go beyond what we discuss here)\footnote{We note that the Ponzano--Regge amplitude itself arises from a discretisation of a $BF$ theory path integral on a given triangulation \cite{BaezBF}, which would include the quantum-mechanical weight $e^{{\rm i}S_{{\rm BF}}}$. This fact seems to be unrelated to the choice of statistical weight $e^{-S[\phi]}$ in the GFT formulation.}. Let us point out that (\ref{GFTpert}) is not defined as a quantum-mechanical path integral but more akin to a statistical partition function; this is the usual definition also used in matrix and tensor models, presumably to allow for a more rigorous definition of the integral in analogy with how functional integrals are defined in Euclidean quantum field theory. In general, little is known about whether (\ref{GFTpert}) converges and the action $S[\phi]$ is typically not bounded from below; one might then alternatively use a path integral with exponent ${\rm i}S[\phi]$, with the understanding that the definition is ill-defined without regularisation and/or analytic continuation (as is almost always the case in quantum field theory). The connection with spin foams would seem to be unchanged by such a revised proposal as we could send $\lambda\rightarrow-{\rm i}\lambda$ while leaving the amplitude for given $C$ unchanged. Notice also that the connection to spin foams does not fix the interpretation of the parameter $\lambda$, which would have to come from elsewhere (e.g., possible phenomenology of GFT models).

(\ref{GFTpert}) is the promise of group field theory: for suitable action $S$, a perturbative expansion in Feynman graphs can be interpreted as a sum over all possible discrete spacetimes weighted by an amplitude that can be interpreted as a discrete quantum gravity path integral (or spin foam amplitude). This statement does indeed hold in quite some generality \cite{Reisenberger}. Naturally, quantum gravity in four dimensions is much more difficult than three, so the construction of suitable $S$ is significantly more involved, but has been done for many spin foam models (see, e.g., \cite{GFTmodels}).\footnote{Four-dimensional models require simplicity constraints to go from topological models to those with local degrees of freedom. How to implement these is an important discussion in the literature, but goes beyond this review which is focused more on structural and conceptual aspects.}  

\begin{figure}[htp]
\includegraphics[scale=.7]{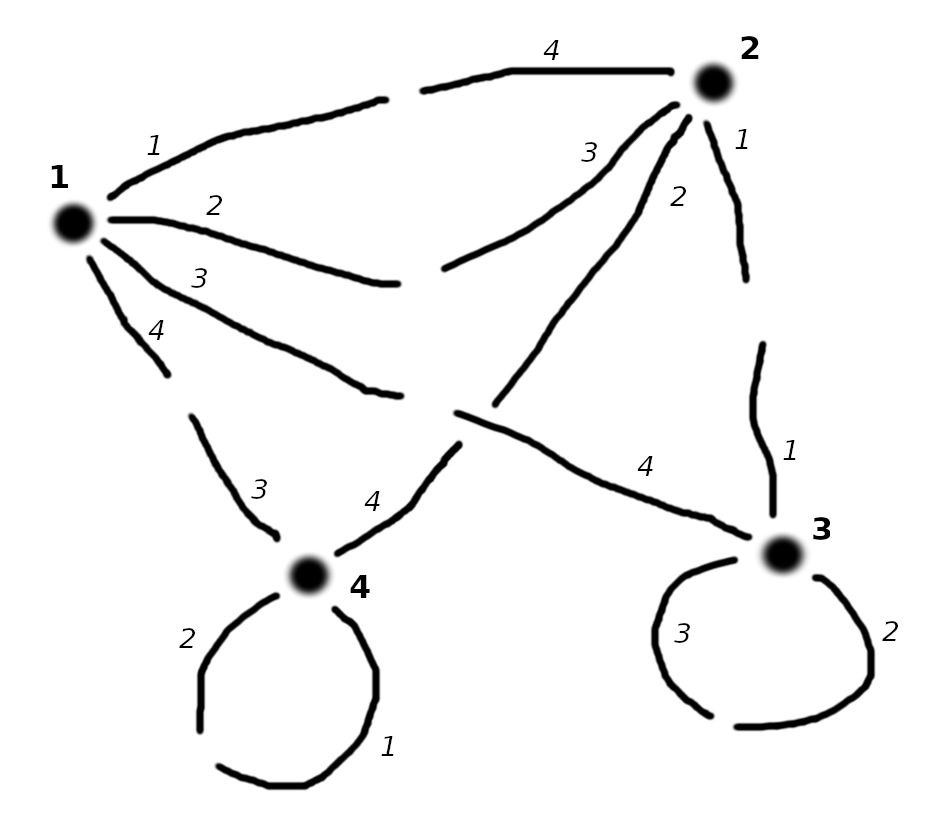}
\caption{A graph with four four-valent vertices; we have split each edge into a ``source'' and a ``target'' part.}
\label{Graphfig}
\end{figure}

Beyond the partition function on its own, one can define GFT observables \cite{GFTfreidel} -- these are observables in the sense of quantum field theory, i.e., functions of $\phi$, not directly related to gravitational observables (which would have to be invariant under suitably defined diffeomorphisms). The most natural observables are associated to $d$-valent spin networks, where $d$ is the number of ${\rm SU}(2)$ arguments in the definition of the field (so $d=3$ for the Boulatov model). A {\em spin network} is an assignment of ${\rm SU}(2)$ irreducible representations $j_e$ to edges $e$ and of intertwiners $\iota_v$ to vertices $v$ of a graph $\Gamma$. Spin networks are ubiquitous in loop quantum gravity, where they define basis states in a canonical quantisation of general relativity \cite{LQGbooks} (to obtain the full basis, vertices with arbitrary valency must be included). In group field theory, for any $d$-valent $\Gamma$ we can define
\ben
\label{Ointegrals}
\mathcal{O}_{\Gamma,\psi}(\phi) = \int {\rm d}^{n\cdot d}g \;\psi((g_{ij})^{-1}g_{i'j'}) \prod_{i=1}^n \phi(g_{ij})\,;
\een
here $i=1,\ldots,n$ labels the vertices of $\Gamma$ and at each vertex we have $d$ incident edges $j=1,\ldots,d$. Each edge starts and ends at a vertex, and is characterised by two pairs $(ij)$ and $(i'j')$. An example with $n=d=4$ is given in figure \ref{Graphfig}. The graph wavefunction $\psi$ in (\ref{Ointegrals}) only depends on the $n\cdot d/2$ combinations $(g_{ij})^{-1}g_{i'j'}$ where $(ij)$ and $(i'j')$ are associated to the same edge. To obtain a spin network state with $(j_e,\iota_v)$, $\psi$ is defined by contracting matrix representations of $(g_{ij})^{-1}g_{i'j'}$ in the representations $j_e$ with intertwiners $\iota_v$ (again see \cite{LQGbooks} for how this works in loop quantum gravity).  

Spin network states form a basis of the space of all possible wavefunctions, which can be characterised as complex-valued functions\footnote{To represent a loop quantum gravity state, $\psi$ would also need to be square-integrable.} on ${\rm SU}(2)^{n\cdot d}$ such that
\ben
\psi(g_{11},\ldots,g_{ij},\ldots,g_{i'j'},\ldots,g_{nd})=\psi(g_{11},\ldots,hg_{ij},\ldots,hg_{i'j'},\ldots,g_{nd})\quad\forall h\in {\rm SU}(2)
\label{symmetry1}
\een
whenever $(ij)$ and $(i'j')$ are assigned to the same edge in $\Gamma$. By choosing $h=(g_{ij})^{-1}$ on the right-hand side of (\ref{symmetry1}), we can see that this reduces to the previous definition. In addition, given that we integrate over group fields $\phi$ normally assumed to satisfy right invariance $\phi(g_{ij})=\phi(g_{ij}h_i)$, and using the fact that we can reorder the product of $n$ fields, we also have
\ben
\label{symmetry2}
\psi(g_{ij})=\psi(g_{ij}h_i)\quad\forall \{h_i\}\in {\rm SU}(2)^n\,,\quad \psi(g_{ij})=\psi(g_{\sigma(i)j})\quad\forall\sigma\in{\mathcal{S}}_n\,.
\een
(These do not strictly need to be imposed on $\psi$ but any observable defined by (\ref{Ointegrals}) can be written in terms of a $\psi$ with those symmetries.)  The first of these symmetries is exactly the local action of ${\rm SU}(2)$ on vertices in loop quantum gravity, and is achieved for spin network states via the contraction with intertwiners. The second one is a discrete symmetry under the symmetric group $\mathcal{S}_n$ corresponding to permutations of vertices, which can be seen as a restriction of wavefunctions or the statement that $\Gamma$ is equivalent to any graph obtained from $\Gamma$ after relabelling of vertices.

The correlation function
\ben
\langle \mathcal{O}_{\Gamma,\psi}(\phi)\mathcal{O}_{\Gamma',\psi'}(\phi)\rangle = \int \mathcal{D}\phi\;\mathcal{O}_{\Gamma,\psi}(\phi)\mathcal{O}_{\Gamma',\psi'}(\phi)\;e^{-S[\phi]}
\een
can be interpreted as a sum over spin foams for which $(\Gamma,\psi)$ and $(\Gamma',\psi')$ appear as boundary states; these are the amplitudes one is often interested in, as they are interpreted as transition amplitudes (or rather, a proposal for a physical inner product) between those asymptotic states. (The precise interpretation of these correlation functions in GFT needs a specification of the assumed operator ordering in the corresponding canonical formalism; we will return to this point below (\ref{SDeqs}).)

{\bf Extending the most basic definitions.} The discussion so far is a summary of the basic idea behind group field theories and their interpretation as a definition (or extension) of spin foam models as a quantum field theory. More recently more sophisticated models have been developed, incorporating further insights that go beyond this identification of individual Feynman amplitudes. We list some motivations for modifying the original definitions: 

\begin{itemize}
\item {\bf Nontrivial kinetic term.} Seen as a quantum field theory, the proposal detailed above is a bare action which requires renormalisation: divergences appearing in the Feynman expansion need to be reabsorbed into a redefinition of the action. In the simplest case of the Boulatov model (\ref{boulatovaction}), one finds that  a nontrivial divergence can only be absorbed if one adds a Laplacian into the kinetic term \cite{Radiative}, i.e.,
\ben
\label{GFTrenormal}
S_0[\phi] = \frac{1}{2}\int{\rm d}^3 g\;\phi^2(g_1,g_2,g_3) \quad\rightarrow\quad S_0^{{\rm ren}}[\phi] = \frac{1}{2}\int{\rm d}^3 g\;\phi(g_1,g_2,g_3)\left(\Delta+m^2\right)\phi(g_1,g_2,g_3)
\een
in terms of the Laplace--Beltrami operator $\Delta$ on ${\rm SU}(2)^3$ and a mass term $m^2$. (Here and in the following we write $S_0[\phi]$ to denote the quadratic part of the action.) Hence, renormalisation forces us into extending the original model. The  correspondence of a group field theory Feynman expansion with spin foam amplitudes is now modified, since the amplitudes will evidently be modified as well. Inclusion of derivative terms into the kinetic term had already been considered previously purely on the basis of reproducing more general discrete quantum gravity amplitudes \cite{Tlas}. We refer to the literature for further details on these developments, but we note that the requirement of renormalisability or even systematic inclusion of radiative corrections necessitates adding derivatives into the kinetic term \cite{renormreview}. Needless to say, this modification also makes group field theories look more like ``ordinary'' quantum field theories. Recent work on renormalisation and phase structure of group field theories such as \cite{LandauGinz} also uses essentially the same form of kinetic term.
\item {\bf Complex field or many fields.} Simple models such as the Boulatov model, based on a real scalar field $\phi$ and an action dictated by combinatorics alone, are known to have various drawbacks. One is the fact that the combinatorial information of a given Feynman graph should be enough to determine the corresponding simplicial complex with its combinatorial structure and topology, but this is not the case in general; another is the absence of the analogue of a $1/N$ expansion, perhaps the most powerful tool for the continuum limit of matrix models \cite{matrixreview}. To rectify these, various generalisations of group field theories have been proposed: {\em coloured group field theory} is based on replacing the single scalar group field by $d+1$ fermionic fields or $d+1$ complex scalar fields \cite{renormreview,colourGFT}. Associating ``colour'' to the fields allows defining a more restricted class of interactions, where only certain combinatorial gluings are possible and resulting Feynman graphs can be interpreted more straightforwardly. In general, considering complex fields can be used to define a new class of interaction terms based on {\em tensor invariants}, in which generally the field $\varphi$ is glued to its conjugate $\bar\varphi$; these can be used to construct renormalisable theories even for a single field \cite{renormreview,tensorGFT}.  
\item {\bf Matter coupling.} In the presentation so far we have focused attention on models for pure vacuum quantum gravity; for more realistic theories we would like to include matter degrees of freedom as well. Here we will mostly focus on scalar matter degrees of freedom and spin foam models including those have only appeared recently \cite{SFmatter}, and to our knowledge have not yet been translated into group field theory language. (For earlier work on including fermionic matter see, e.g., \cite{vierbein}.) However, insights into what modifications are needed at the kinematical level can be obtained by considering loop quantum gravity spin networks for gravity and matter \cite{LQGbooks}. In loop quantum gravity (or a more general lattice regularisation), matter degrees of freedom appear as additional data on a graph; a discretised scalar field (or $0$-form) naturally sits on vertices. The interpretation (\ref{Ointegrals}) of a group field as associated to a spin network vertex then suggests extending the domain of $\phi$: to couple $D$ scalar matter fields we should now consider a field (here chosen to be real for simplicity)
\ben
\phi: G^d\times\mathbb{R}^D \rightarrow \mathbb{R}\,,\quad \phi(g_i,\chi^a)=\phi(g_ih,\chi^a)\quad\forall h\in G\,,\een
with the last equation meaning that local $G$ gauge transformations should leave the scalar field arguments invariant, and we have now generalised from ${\rm SU}(2)$ to a general compact gauge group $G$. While this fixes the kinematical structure of an extended group field theory, the choice of action needs to be motivated in other ways. Various proposals have been made, including \cite{GFTscalar} which uses an identification of Feynman amplitudes with discrete (gravity and matter) path integrals; this model includes a kinetic term nonlocal in $\chi$.
\end{itemize}

It should be clear from this overview (which is by no means complete -- for instance, we will not discuss the important extension to non-compact groups) that a large number of models have been proposed based on different motivations. We are looking for a model that not only has a clear interpretation of four-dimensional quantum gravity plus matter, but is also well-behaved as a quantum field theory and provides a perturbative expansion that is under control in a meaningful sense. Without such a theory, which features are preferred or seen as essential in the construction of models can be seen to an extent as a matter of choosing priorities. All existing literature on Hilbert space formalisms for group field theory assumes a nontrivial kinetic term; there are discussions of real and complex group fields; matter fields have been included by extending the domain of the group field, with different assumptions on the action.

\section{How to get a Hilbert space I: algebraic approach}

So far, we introduced the group field theory formalism via a Lagrangian and associated path integral or partition function (where we use the term partition function to indicate that the weight inside the integral is $e^{-S}$ rather than $e^{{\rm i}S}$). With no obvious notion of time variable, it is not straightforward to develop a corresponding Hamiltonian formulation that would be the basis for canonical quantisation.

\subsection{Main ideas and connection to loop quantum gravity}

The proposal of \cite{Oriti2q} is to construct a kinematical Hilbert space based on the interpretation of group field theory states (or observables) (\ref{Ointegrals}) as spin network states in loop quantum gravity. The wavefunction $\psi$ appearing in (\ref{Ointegrals}) looks very much like an $n$-particle wavefunction in many-body quantum mechanics, where each particle lives on a $G^d$. The ``$n$-particle'' wavefunction on $G^{n\cdot d}$ has additional symmetries (\ref{symmetry1}) needed for a spin network interpretation and (\ref{symmetry2}) coming from symmetries of the group field and commutativity of real-valued fields $\phi(g_{ij})$.

Quantum field theory textbooks \cite{giftedama} tell us how to pass from $n$-particle wavefunctions with bosonic permutation symmetries to a ``second quantised'' picture in terms of annihilation and creation operators. Namely, we can define such operators to satisfy
\ben
[ \hat{a}(g_j), \hat{a}^\dagger(g'_j) ] = I(g_j(g'_j)^{-1}) := \int_{G} {\rm d}h\;\delta^{(4)}(g_jh(g'_j)^{-1}) 
\label{firstCCR}
\een
such that $\hat{a}(g_j)=\hat{a}(g_jh)$ and $\hat{a}^\dagger(g_j)=\hat{a}^\dagger(g_jh)$ for all $h \in G$, and assume that there is a Fock vacuum $|0\rangle$ satisfying $\hat{a}(g_i)|0\rangle = 0$. More general states in the Fock space are then defined from repeated action of creation operators on $|0\rangle$ and the inner product on this Hilbert space is induced by (\ref{firstCCR}). For instance, for single-particle states $|g_i\rangle := \hat{a}^\dagger(g_i)|0\rangle$ we have
\ben
\langle g_i|g'_i\rangle = \langle 0| \hat{a}(g_i) \hat{a}^\dagger(g'_i)|0 \rangle = I(g_i(g'_i)^{-1})\,.
\een
In this Hilbert space one could now define
\ben
|\psi\rangle = \int {\rm d}^{n\cdot d}g \;\psi((g_{ij})^{-1}g_{i'j'}) \prod_{i=1}^n \hat{a}^\dagger(g_{ij})|0\rangle
\label{GFTstate1}
\een
as the quantum state representing  (\ref{Ointegrals}) in a Hilbert (Fock) space.  Notice that $\hat{a}$ and $\hat{a}^\dagger$ have the same symmetries as the group field $\phi$ appearing above, and (\ref{firstCCR}) is defined to be compatible with those symmetries. Hence, each observable $\mathcal{O}_{\Gamma,\psi}(\phi)$ defined via (\ref{Ointegrals}) can be represented as a state (\ref{GFTstate1}) by replacing all fields $\phi(g_{ij})$ by creation operators $\hat{a}^\dagger(g_{ij})$ acting on $|0\rangle$. In this sense, one can view the whole construction as a second quantisation representation of loop quantum gravity spin networks. The key new idea is now to identify $\hat{a}(g_j)$ and $\hat{a}^\dagger(g_j)$ with a {\em complex} group field and its conjugate \cite{Oriti2q}: $\hat\varphi(g_j)\equiv\hat{a}(g_j)$ and $\hat\varphi^\dagger(g_j)\equiv\hat{a}^\dagger(g_j)$. Given this starting point of assuming an algebra of field operators and deriving everything from that, the approach has been called {\em algebraic approach} to obtaining a GFT Hilbert space \cite{GaussianGFT}. We must now assume a complex field $\varphi$ even though (\ref{Ointegrals}) is often defined for a real $\phi$.

The construction mimics the usual introduction of field operators from creation and annihilation operators in non-relativistic many-body quantum physics \cite{giftedama}. There, the operators satisfy first order differential equations in time (the second quantised Schr\"odinger equation). In relativistic field theories, on the other hand, the field equations are second order and the conjugate momentum to a field is not given by its complex or Hermitian conjugate, but by a time derivative of the field. Relativistic field theories are well-defined for real fields whereas the non-relativistic version needs a complex field because the field and its conjugate do not commute, as in (\ref{firstCCR}). Defining the commutation relations for group field theory based on analogy with non-relativistic quantum field theory seems to be an assumption that cannot directly be motivated from the structure of dynamical equations, but could be argued to be the only option if no background time is available. As for spin networks in loop quantum gravity, the definitions are meant to be kinematical with dynamical information to be imposed at a later stage \cite{LQGbooks}.

As we have illustrated in the beginning of this section, the definitions made so far already fix the inner product between two states of the form (\ref{GFTstate1}) via the algebra of $\hat\varphi(g_j)$ and $\hat\varphi^\dagger(g_j)$ and the normalisation condition $\langle 0|0\rangle=1$. Recall that the wavefunction is assumed to satisfy  (cf.~(\ref{symmetry1}))
\ben
\psi(g_{11},\ldots,g_{ij},\ldots,g_{i'j'},\ldots,g_{nd})=\psi(g_{11},\ldots,hg_{ij},\ldots,hg_{i'j'},\ldots,g_{nd})\quad\forall h\in {\rm SU}(2)
\label{symmetryagain}
\een
if the state is associated with a spin network in which $(ij)$ and $(i'j')$ are identified (``glued together'') as the same edge. For two such states with the {\em same} symmetries, and hence associated with the same graph, the resulting inner product is the one expected from loop quantum gravity. However, two states associated to different graphs with the same number of vertices are in general not orthogonal whereas they would be in loop quantum gravity. This  point is already discussed in \cite{Oriti2q}.

In general, since the wavefunction is required to have additional symmetries, only rather specific Fock states have a spin network interpretation. This identification of a subspace in the Fock space is a little similar, although not quite equivalent, to the passage from gauge-variant kinematical states to gauge-invariant spin networks in loop quantum gravity. Given a state $|\psi\rangle$, one may be able to reconstruct its associated graph by checking for symmetries of the form (\ref{symmetryagain}), though there are counterexamples -- for instance, a constant $\psi$ would be compatible with any possible graph.

Up to this point, the GFT Hilbert space has been constructed as an abstract Fock space with no direct input from the GFT (classical or quantum) dynamics as defined by the action. We also saw that the requirement for GFT states to be interpreted in terms of loop quantum gravity spin networks leads to gluing conditions (\ref{symmetryagain}) which have to be imposed separately. A crucial question for this approach to quantisation is then how to identify states that are ``allowed'' within an initial ``too large'' state space.  Following the philosophy of constraint quantisation {\em \`a la} Dirac, we may say that we are looking for {\em physical} states in a {\em kinematical} Hilbert space. The construction so far does not arise from a Dirac-type analysis (but see (\ref{KGaction2}) and below), nor are there constraints related to any gauge symmetries of the GFT, as we will also discuss below. However, given that the classical GFT field equations 
\ben
\frac{\delta S[\varphi,\bar\varphi]}{\delta\varphi(g_i)} = \frac{\delta S[\varphi,\bar\varphi]}{\delta\bar\varphi(g_i)} = 0
\label{classicaleom}
\een
single out particular classical field configurations as physical, one might expect there to be a connection between the space of physical states in the quantum theory and the solutions of the classical field equations, as there is in standard quantum field theory.

For a real action $S$, these two equations are the same up to complex conjugation, so one can restrict to solving one of them. The question is now how to represent (\ref{classicaleom}) on the kinematical Fock space, and two main ideas have been proposed. The first is to view (\ref{classicaleom}) as constraints and demand that their operator version annihilates physical states \cite{GFTcosmorel},
\ben
\label{constraint1}
\widehat{\frac{\delta S[\varphi,\bar\varphi]}{\delta\bar\varphi(g_i)}} |\Psi\rangle = 0\,.
\een
Now the requirement that both equations of motion should annihilate physical states will lead to further constraints, since the two operators representing the two equations (\ref{classicaleom}) do not commute. Indeed, in the simplest case of a free theory, where $\widehat{\delta S/\delta\varphi}$ is linear in $\hat\varphi^\dagger$ and $\widehat{\delta S/\delta\bar\varphi}$ is linear in $\hat\varphi$, and with $[ \hat{\varphi}(g_j), \hat{\varphi}^\dagger(g'_j) ] = I(g_j(g'_j)^{-1}) $, the commutator of the two constraints is proportional to the identity, so that the two constraint equations cannot be solved simultaneously. One may be content with assuming that one equation holds strongly and the conjugate one ``weakly'' as an expectation value. In the next section we will describe another approach focused on replacing (\ref{constraint1}) by a single Hermitian constraint operator whose classical limit is equivalent to solving the equations of motion. In either approach, for an interacting GFT the explicit construction of the solution space to (\ref{constraint1}) will be generally impossible, as it is in standard field theory. The technical discussions below will hence refer mostly to the approximation of a free GFT with an action quadratic in the fields. At least in this case, it would seem reasonable to expect that any proposed construction can be implemented and understood completely. Then, while one may be able to construct the subspace of kinematical states satisfying (\ref{constraint1}), these are generally found to be non-normalisable (see below).

An alternative, proposed already in \cite{Oriti2q,GFTcondensate}, is to use Schwinger--Dyson equations derived from a path integral formulation. Indeed, a standard field theory argument is that
\ben
0 = \int \mathcal{D}\varphi\;\mathcal{D}\bar\varphi\,\frac{\delta}{\delta\bar\varphi(g_i)}\left(\mathcal{O}[\varphi,\bar\varphi]e^{{\rm i}S[\varphi,\bar\varphi]}\right) = \left\langle \frac{\delta\mathcal{O}[\varphi,\bar\varphi]}{\delta\bar\varphi(g_i)} + {\rm i}\mathcal{O}[\varphi,\bar\varphi] \frac{\delta S[\varphi,\bar\varphi]}{\delta\bar\varphi(g_i)}\right\rangle
\een
where $\mathcal{O}[\varphi,\bar\varphi]$ can be any polynomial in the fields. The assumption is now that any such constraint between correlation functions derived in the path integral language can be expressed as operator expectation values in the quantisation proposed above, i.e., assuming the particular commutators between $\hat\varphi(g_j)$ and $\hat{\varphi}^\dagger(g'_j)$, and in any physical state. Satisfying {\em all} these equations (for all possible $\mathcal{O}[\varphi,\bar\varphi]$) is then taken as a definition of what constitutes a physical state, seen as a nonperturbative vacuum of the theory.

In general, the first two Schwinger--Dyson equations (for $\mathcal{O}=1$, $\mathcal{O}=\bar\varphi(g'_i)$) would be
\ben
\left\langle \frac{\delta S[\varphi,\bar\varphi]}{\delta\bar\varphi(g_i)}\right\rangle =0\,,\quad \left\langle \bar\varphi(g_i')\frac{\delta S[\varphi,\bar\varphi]}{\delta\bar\varphi(g_i)}\right\rangle = -{\rm i}\, I(g_j(g'_j)^{-1})\,.
\label{SDeqs}
\een
If we restrict to a free theory for simplicity, these equations characterise the one-point and two-point functions: the first one says that the expectation value of the field operator satisfies the classical equation of motion, which is certainly the case in standard quantum field theory; the second one, in standard (free) quantum field theory, characterises the Feynman propagator or time-ordered two-point function evaluated in the vacuum state (which is unique in the free theory). In an operator formalism the equation can in general only hold for a specific operator ordering, and the distributional (contact) term can be seen as arising from a commutator. In standard quantum field theory the resulting Schwinger--Dyson equations hold for time ordering \cite{PeskinSchroeder} which has no immediate analogue in GFT. Notice that (\ref{constraint1}) would be incompatible with the second equation in (\ref{SDeqs}) if we adopt an operator ordering in which $\widehat{\delta S/\delta\bar\varphi}$ ends up on the right.

No systematic analysis of Schwinger--Dyson equations beyond the simplest cases has been attempted in the algebraic approach, and the required assumptions about operator ordering and, in particular, the commutators of $\hat\varphi(g_i)$ and $\hat\varphi^\dagger(g_j)$, have not been independently justified. There is hence so far no concrete definition of the physical Hilbert space, even for free GFT.  

On general grounds, if we demand that the space of physical states has at least the structure of a vector space, it should arise as the kernel of one or multiple operators acting on kinematical states, i.e., there should be constraint equations. The most natural candidate seems to be (\ref{constraint1}) (or its conjugate), but a different idea was proposed in \cite{Oriti2q}, where an extended GFT action is itself seen as representing a constraint. This second idea is naturally associated to an interpretation of the GFT functional integral as a statistical partition function, so that the action $S$ appears as the analogue of a Hamiltonian or free energy. This second proposal seems to not have been investigated much in follow-up work and it would suggest a different conceptual view on GFT as a theory fundamentally based on statistical ensembles rather than pure states. In either case, if a constraint operator has a continuous spectrum\footnote{For a GFT defined only on a compact group (i.e., in particular without matter coupling), there may be a constraint operator with discrete spectrum, e.g., involving the number operator associated to a discrete set of Peter--Weyl representation modes. To the best of our knowledge this case has not been discussed in the literature.}, its kernel will not define a subspace of the kinematical Hilbert space, since states defined in this way are not normalisable. This is a common problem in constraint quantisation which can be addressed by redefining the inner product, as we do below. Very few discussions of this non-normalisability issue can be found in the literature; the analysis of \cite{GaussianGFT} found that the Fock vacuum seems to be the only well-defined physical state that lives in the original kinematical Hilbert space.

\subsection{Building foundations for the algebraic approach}

We have seen that the algebraic approach defines a Hilbert space formalism for GFT whose states can be interpreted as kinematical spin network states; however, it requires additional assumptions on the basic algebra of field operators and there is no clear characterisation of physical states in the Fock space. A few attempts have been made to provide firmer foundations for this approach, and relate it to established canonical quantisation techniques.

An extensive discussion of possible connections between the functional integral and a Hilbert space formalism for GFT can be found in  \cite{KegelesPhD}, mentioning some of the points we discuss in this review from similar or different viewpoints. The discussion culminates in a novel proposal for defining a symplectic structure on the space of group field configurations from the GFT action; such a construction allows interpreting the space of field configurations as a phase space and directly leads to an algebra of field operators, alleviating the concern that commutators of field operators have been simply postulated in the algebraic approach.

The construction uses many ingredients from algebraic quantum field theory, where the focus shifts away from the Hilbert space towards an algebra of observables; here we attempt to summarise the main results. First, assume a GFT for a real field $\phi$ (here on ${\rm SU}(2)^d$) and expand the action to second order around a classical (background) solution $\phi_0$,
\ben
S[\phi] = S[\phi_0+\xi] = S[\phi_0] + \frac{1}{2}\int {\rm d}^d g\;{\rm d}^d h\;\xi(g_i)\mathcal{K}\xi(h_i) + \ldots
\een
where $\mathcal{K}$ is in general a differential operator arising from the second variation of the original action $S[\phi]$. The background solution $\phi_0$ can be zero or a nontrivial solution of the interacting theory, and this is associated with a perturbative expansion of the theory around different ``phases''. Dropping the higher-order terms, we obtain a Gaussian functional integral (defined in the Euclidean/statistical sense with integrand $e^{-S[\phi]}$) that can be evaluated exactly, and all correlation functions are defined in terms of the two-point function
\ben
\langle\xi(g_i)\xi(h_i)\rangle = \mathcal{G}(g_i,h_i)
\een
where $\mathcal{G}(g_i,h_i)$ is a suitably chosen Green's function for the differential operator $\mathcal{K}$. The Green's function induces an inner product on complex functions on ${\rm SU}(2)^d$ via
\ben
\langle f,f'\rangle_{\mathcal{G}} = \int {\rm d}^d g\;\bar{f}(g_i)\mathcal{G}(g_i,h_i)f'(h_i)
\een
whose imaginary part defines a symplectic form on the space of complex functions. This symplectic form allows the construction of a Weyl algebra of observables and, after a choice of Fock algebraic state (defined as a functional on the space of observables), the Weyl algebra can be written as generated by a self-adjoint operator $\hat\Phi_F(f)$ for each $f$. Suitable linear combinations then lead to the definition of ladder operators $\hat{A}(f)$ and $\hat{A}^\dagger(f')$ satisfying the canonical commutation relations of the algebraic approach, such that the chosen algebraic state becomes a vacuum annihilated by $\hat{A}(f)$, and we obtain the desired Fock representation. The construction provides a map between the field  $\xi$ appearing in a functional integral and the creation and annihilation operators encoded in $\hat{A}(f)$ and $\hat{A}^\dagger(f')$; the map is nontrivial and depends on the operator $\mathcal{K}$, and will not simply identify field operators with the ladder operators as proposed above.

The result of the work of \cite{KegelesPhD} is hence a Hilbert (Fock) space formalism for a general GFT, expanded perturbatively around a background solution, such that correlation functions of field operators agree with functional integral results. The approach does not require a notion of time,  but it does require knowledge of the action $S$ and specifically a Green's function for the operator $\mathcal{K}$ which will be available for simple enough theories but, e.g., presumably not in the most general nonlocal case we mentioned earlier. It is also not yet understood whether there is a subspace of physical states characterised by some set of constraints.\footnote{Given the non-normalisability issue of physical states whenever constraints are imposed, a possible suggestion might be that there are no constraints at all, but this seems to be at odds with the classical limit in which only some field configurations are physical (solve the equations of motion). See also footnote \ref{finalfootnote}.}

A rather different approach proposed in \cite{frozenform} is to redefine the action in such a way that classical solutions correspond to the ones of the original GFT action, but that a conventional canonical quantisation is possible. This requires introducing a notion of time parameter from which a phase space structure can be derived, but given that the initial GFT has no notion of time, this must be a fiducial structure that the theory does not actually depend on.

The idea can be illustrated in the example of a free complex Klein--Gordon scalar field, with action (defined in momentum space)
\ben
\label{KGaction}
 S[\Phi,\bar\Phi] = -\int \frac{{\rm d}^D p}{(2\pi)^D} \,\bar\Phi(p^\mu)\,(p^2+m^2)\Phi(p^\mu)
\een
and Klein--Gordon equation $(p^2+m^2)\Phi(p^\mu)=(p^2+m^2)\bar\Phi(p^\mu)=0$. As we mentioned after (\ref{constraint1}), we really have two constraints that are conjugates of each other, even though they are classically equivalent; but here we can replace them by the equivalent constraint $(p^2+m^2)|\Phi|^2=0$. Such a constraint can be imposed through the usual Lagrange multiplier method. Finally, we would like to identify $\Phi$ and $\bar\Phi$ as canonically conjugate quantities on a kinematical Hilbert space. All this can be achieved by proposing a ``frozen'' action
\ben
\label{KGaction2}
S[\Phi,\bar\Phi,N] = \int \frac{{\rm d}^D p}{(2\pi)^D} \,{\rm d}\tau\left[\frac{{\rm i}}{2}\left(\bar\Phi(p^\mu,\tau)\dot\Phi(p^\mu,\tau)-\Phi(p^\mu,\tau)\dot{\bar\Phi}(p^\mu,\tau)\right)+N(p^\mu,\tau)(p^2+m^2)|\Phi(p^\mu,\tau)|^2\right]
\een
where we have introduced a ``lapse'' field $N$, and all fields now have an additional dependence on a fiducial time parameter $\tau$; $\dot{}$ denotes derivative with respect to $\tau$. The equations of motion now imply that $(p^2+m^2)|\Phi(p^\mu,\tau)|^2=0$, which should be seen as a primary constraint, and that $\Phi$ and $\bar\Phi$ are actually independent of $\tau$. We also see that $\Phi$ and $\bar\Phi$ are canonically conjugate. Moreover, the constraint is naturally represented as a Hermitian operator in the quantum theory.

The theory can now be quantised along the lines of a constraint/Dirac quantisation \cite{DiracQ}, suitably generalised to field theory. One can define a kinematical Fock space, generated by canonically conjugate creation operators $\hat\Phi^\dagger(p^\mu)$ and annihilation operators $\hat\Phi(p^\mu)$. This is then the analogue of the kinematical Fock space proposed in the algebraic approach for GFT, and indeed can be defined for free GFT along the exact same lines. For the Klein--Gordon field, the constraint becomes $(p^2+m^2)\hat\Phi^\dagger\hat\Phi=0$ where $\hat\Phi^\dagger\hat\Phi$ is the number density in a given mode; hence only particles satisfying $p^2+m^2=0$ correspond to physical states. For given spatial momentum $\vec{p}$, there are two solutions $p^0=\pm\sqrt{\vec{p}^2+m^2}$, naturally identified as particle and antiparticle. The kinematical Hilbert space contains no normalisable physical states (apart from the vacuum) and so one has to define a different physical inner product, which can be done via {\em group averaging}, a standard procedure employed in loop quantum cosmology \cite{LQC} and loop quantum gravity \cite{LQGbooks} (see also \cite{HoehnRelP} for a concrete example in a relativistic particle model). Again, there are some new subtleties in a second quantised picture, and the idea of \cite{frozenform} is to define a generalised projection of operators defined on the kinematical Fock space to well-defined operators on the physical Fock space. Such a projection does not exist for all operators but only the ones that commute with the constraint at least weakly, again in accordance with the Dirac programme. In this way, it is possible to systematically construct a physical Hilbert space and identify operators that can be interpreted as observables, questions that are not discussed in most presentations of the algebraic approach.

The specific idea of replacing the action (\ref{KGaction}) by an extended action (\ref{KGaction2}) works well for actions quadratic in the fields, but it has no obvious analogue once nonlinearities are included, even before considering the fact that a more complicated constraint would need to be solved. Hence, extending these ideas to interacting theories is far from straightforward. Of course, even in standard quantum field theory we construct the Fock space for free theories and then work in perturbation theory, so perhaps this is not something one could expect. One could also argue that the specific proposal (\ref{KGaction2}) is a bit artificial, as it is designed to make $\Phi$ and $\bar\Phi$ conjugate. To an extent, the resulting phase space structure is still postulated, although more clearly motivated. No functional integral formulation for actions of the type (\ref{KGaction2}) has been considered which would link back to spin foams.

\section{How to get a Hilbert space II: deparametrised approach}

As we have already discussed, the construction of a Hilbert space formalism for GFT is complicated by the fact that there is, in general, no notion of time that could be used to construct a classical phase space as the basis for canonical quantisation in the usual way ($\{\cdot,\cdot\}\rightarrow -\frac{{\rm i}}{\hbar}[\cdot,\cdot]$). In some models, however, one can identify clocks based on analogy with other approaches to quantum gravity and quantum cosmology. The prime candidate is a free massless scalar matter field as used in loop quantum cosmology \cite{LQC}, and such a matter degree of freedom was indeed used to define the first notion of relational dynamics in GFT \cite{GFTcosmorel}. In this first proposal, the idea was to work on the kinematical Hilbert space of the algebraic approach and merely define different operators similar to relational observables; later on, it was realised that one could also define a different quantisation based on using this clock as a notion of time throughout \cite{relHamilt}. The process is conceptually similar to {\em deparametrisation}, i.e., gauge-fixing a reparametrisation symmetry by identifying a suitable clock classically before quantisation, although this is usually more an analogy given that GFT does not have a Hamiltonian formulation and is  not a constrained system in a clear sense (however, it can be written in this form, as we will discuss below).

We can define this deparametrised approach for a group field on $G^d\times \mathbb{R}$ where often the specific choice $G^d={\rm SU}(2)^4$ is made, although the formalism is more general. Assuming that $G$ is compact, there exists an expansion of the field $\phi(g_i,\chi)$ into representation modes
\ben
\phi(g_i,\chi) = \sum_J \phi_J(\chi)D_J(g_i)
\een
where $J$ is a multi-index encoding a sufficient number of discrete labels, and $D_J(g_i)$ represent a combination of (entries of) representation matrices. For instance, for $G^d={\rm SU}(2)^4$ (see, e.g., \cite{geneff})
\ben
\phi(g_i,\chi) = \sum_{j_I\in{\rm Irrep}}\sum_{m_I,n_I=-j_I}^{j_I}\sum_\iota \phi_{m_I}^{j_I,\iota}(\chi) \,\mathcal{I}_{n_I}^{j_I,\iota}\prod_{\alpha=1}^4\sqrt{2j_\alpha+1}D^{j_\alpha}_{m_\alpha n_\alpha}(g_\alpha)
\een
where $j_I$ are irreducible ${\rm SU}(2)$ representations, $m_I,n_I$ are magnetic indices, $\iota$ are ${\rm SU}(2)$ intertwiners and $D^j_{mn}(g)$ are Wigner $D$-matrices. These are the familiar objects appearing also when constructing spin network states in loop quantum gravity. The canonical quantisation does not strongly depend on these details; only a general expansion into discrete representation modes is needed, which exists due to the Peter--Weyl theorem. A subtle point is that, even assuming a real field $\phi(g_i,\chi)$, the coefficients $\phi_J(\chi)$ are complex but subject to reality conditions.

The starting point is then a quadratic action
\ben
\label{GFTquadrat}
 S_0[\phi] = \frac{1}{2}\sum_J \int {\rm d}\chi\;\bar\phi_J(\chi)\left(\mathcal{K}^{(0)}_J+\mathcal{K}^{(2)}_J\partial_\chi^2\right)\phi_J(\chi)
\een
where $\mathcal{K}^{(0)}_J$ and $\mathcal{K}^{(2)}_J$ are $J$-dependent couplings. The form of the action is rather general, only assuming a kinetic term diagonal in $J$ (obtained if, in terms of $g_i$ variables, the quadratic action only depends on derivatives and not explicitly on $g_i$) as well as shift symmetry $\chi\rightarrow\chi+\chi_0$ and symmetry under parity $\chi\rightarrow-\chi$. The latter two are symmetries of a free massless scalar field and hence it is naturally to impose them on the GFT \cite{GFTcosmorel}. Terms with higher than second derivatives could in principle be present but are usually assumed to be either very small or altogether absent. Indeed, recall that renormalisation even of a trivial kinetic term induces a Laplacian in the group variables (cf.~(\ref{GFTrenormal})) which would lead to an action of the form (\ref{GFTquadrat}), but no higher derivatives are needed from a theoretical point of view; (\ref{GFTquadrat}) is also the quadratic part of the action often adopted in field theoretic studies of GFT \cite{LandauGinz,renormreview}.

Canonical quantisation of (\ref{GFTquadrat}) now proceeds immediately once we accept the interpretation of $\chi$ as a time coordinate.\footnote{Again we want to stress that this is because a {\em classical} Hamiltonian description for GFT is most straightforwardly defined once we have a candidate clock. One could indeed take the view that many conceptual and technical questions arising in our discussion of different quantisations here really arise in classical GFT already.} First, by a simple linear redefinition of the field modes $\phi_J(\chi)$ one can recast the action in terms of explicitly real modes $\phi_J(\chi)$, now without reality conditions \cite{PageWootters}. One then performs an integration by parts to obtain
\ben
\label{GFTquadrat2}
 S_0[\phi] = \frac{1}{2}\sum_J \int {\rm d}\chi\left(\mathcal{K}^{(0)}_J\phi_J^2(\chi)-\mathcal{K}^{(2)}_J(\partial_\chi\phi_J(\chi))^2\right)\,.
\een
For each $J$, this is now a simple particle action in terms of a configuration variable $\phi_J$ and its velocity $\partial_\chi\phi_J$, and the quantisation is entirely straightforward: each $J$ mode is either a harmonic oscillator or an upside-down harmonic oscillator, depending on the signs of $\mathcal{K}^{(0)}_J$ and $\mathcal{K}^{(2)}_J$. (There is also the fine-tuned case $\mathcal{K}^{(0)}_J=0$ which corresponds to a free particle \cite{Condensate}.) The Hamiltonian for each $J$ can then be expressed in terms of ladder operators $\hat{a}^\dagger_J$ and $\hat{a}_J^\dagger$ as
\ben
\hat{H}_J = -{\rm sgn}(\mathcal{K}^{(0)}_J)\omega_J\left(\hat{a}^\dagger_J \hat{a}_J + \frac{1}{2}\right) 
\een
if the signs of $\mathcal{K}^{(0)}_J$ and $\mathcal{K}^{(2)}$ are the same and
\ben
\hat{H}_J = -{\rm sgn}(\mathcal{K}^{(0)}_J)\frac{\omega_J}{2}\left(\hat{a}^\dagger_J \hat{a}^\dagger_J + \hat{a}_J \hat{a}_J\right) 
\een
if they are different, with $\omega_J=\sqrt{|\mathcal{K}^{(0)}_J/\mathcal{K}^{(2)}_J|}$. These are indeed the known expressions for a harmonic oscillator and upside-down harmonic oscillator, with the latter Hamiltonian also called a ``squeezing operator'' given that its action generates squeezed states, e.g., in quantum optics. To be concrete, and working in the Heisenberg picture, one can define the linear combinations
\ben
\hat{\alpha}_J = \sqrt{\frac{\Omega_J}{2}}\,\hat\phi_J +  \frac{{\rm i}}{\sqrt{2\Omega_J}}\,\hat\pi_J\,, \quad \hat{\alpha}^\dagger_J = \sqrt{\frac{\Omega_J}{2}}\,\hat\phi_J -  \frac{{\rm i}}{\sqrt{2\Omega_J}}\,\hat\pi_J\,,
\een
where $\hat\pi_J=-\mathcal{K}^{(2)}_J\partial_\chi\hat\phi_J$ is the canonical momentum conjugate to $\hat\phi_J$ and $\Omega_J=\sqrt{|\mathcal{K}^{(0)}_J/\mathcal{K}^{(2)}_J|}$. In the Heisenberg picture, $\hat{\alpha}_J$ and $\hat{\alpha}^\dagger_J$ are $\chi$-dependent with {\em equal-time} commutation relation
\ben
[\hat\alpha_J(\chi),\hat\alpha^\dagger_{J'}(\chi)]=\delta_{J,J'}\,,
\een
and they define ladder operators $\hat{a}_J=\hat{\alpha}_J(0)$ and $\hat{a}^\dagger_J=\hat{\alpha}_J^\dagger(0)$ which, in the usual way, induce a Fock space via action of $\hat{a}^\dagger_J$ on a vacuum $|0\rangle$ annihilated by all $\hat{a}_J$.

Formally, the resulting Hilbert space can again be interpreted in terms of spin network states, as we discussed in the algebraic approach earlier. $\hat{a}^\dagger_J$ can be seen as creating an open vertex with four links and data determined by $J$, and an extended state could be defined by converting (\ref{GFTstate1}) into the representation basis, or alternatively defining $\hat{a}^\dagger(g_i)=\sum_J\hat{a}^\dagger_J D_J(g_I)$. Notice however the change in perspective; the Fock space is now already a Hilbert space of physical states, with usual time evolution with respect to scalar matter time $\chi$. In the Heisenberg picture, states appear ``timeless'' as they did in (\ref{GFTstate1}), but here this is because we are looking at them at $\chi=0$ while the observables evolve. In the Schr\"odinger picture, states acquire a $\chi$ dependence determined by a Schr\"odinger equation \cite{relHamilt}. The interpretation of these states is hence analogous to loop quantum gravity models in which constraints have been solved via deparametrisation, e.g., by using dust matter or scalar fields, so that there is ``true'' evolution with respect to a matter clock (see, e.g., \cite{DepLQC}). This is why this approach to GFT is usually interpreted in terms of deparametrisation. 

We have focused on the quadratic part of the GFT action here; for this case the dynamics can be solved exactly, in the sense that one can give solutions to the Heisenberg equations for $\hat{a}_J(\chi)$ and $\hat{a}^\dagger_J(\chi)$ and hence the time dependence of any operator in the theory is known. This simple case is enough, e.g., to derive a satisfactory cosmology \cite{geneff}. Interactions can be added to the action; they will not modify the definition of the ladder operators or Fock space, but make the Hamiltonian more complicated and prevent an analytical exact solution. One might try to apply perturbation theory in regimes where the interaction terms are small, but in the most studied cases this approximation breaks down quickly and only full numerics can be used (see again, e.g., \cite{geneff}), as one would expect from a more complicated interacting quantum system. Suitable parameter choices can yield interactions that stabilise the instabilities coming from the quadratic part, and give a new non-perturbative ``condensate'' ground state \cite{Condensate}. It should be said that explicit studies of interacting GFT models have only been done for toy models which include a single $J$ mode.

The deparametrised approach relies on being able to write the quadratic part of the GFT action in the simple form (\ref{GFTquadrat2}), where the role of $\chi$ as a clock is manifest and only fields and first derivatives appear. In theories in which the kinetic term includes higher than second derivatives, more involved methods would need to be applied, and in the case of a theory nonlocal in $\chi$ a canonical quantisation is presumably impossible (one would not even expect a classical initial value formulation). As we mentioned above, there is no clear necessity for including such higher derivatives or nonlocalities, but constructions based on Feynman amplitudes can lead to a nonlocal GFT \cite{GFTscalar}. 

One could also apply the criticisms usually levelled against deparametrised approaches \cite{Isham}, as is done, e.g., in the conceptual discussion of \cite{CPSpaper}. The deparametrised approach to GFT is then classed as {\em tempus ante quantum}: one has selected a degree of freedom (the matter variable $\chi$) as a clock before quantisation, treating it as effectively classical, and such a process would in general break the expected covariance of a quantum theory under time reparametrisations. In simpler terms, this is the worry of how to preserve gauge invariance if a gauge-fixed version of a theory is quantised. Again, such discussions are imported from long-standing discussions in canonical quantum gravity, but is not entirely obvious how they relate to GFT, whose action does not have diffeomorphism symmetry or other gauge symmetries that can be fixed.\footnote{See \cite{GFTdiffeos} for some work identifying symmetries of certain GFT models that correspond to diffeomorphisms. These appear as global, not local (gauge) symmetries; see also \cite{Regensburg} for a more general conceptual discussion of this point.\label{diffeofootnote}} One way to address such concerns is again to reformulate the GFT action to give it an explicit reparametrisation symmetry, which can then be discussed as a gauge symmetry, as we will see now.

\section{Parametrised--deparametrised approach \`a la Page--Wootters}

Starting again with the quadratic action defined in (\ref{GFTquadrat2}),
\ben
 S_0[\phi] = \frac{1}{2}\sum_J \int {\rm d}\chi\left(\mathcal{K}^{(0)}_J\phi_J^2(\chi)-\mathcal{K}^{(2)}_J(\partial_\chi\phi_J(\chi))^2\right)\,.
\een
one can go through the standard process of {\em parametrisation} by promoting $\chi$ to a function of some parameter $\tau$, so that $\phi_J(\chi)=\phi_J(\chi(\tau))$ is then also thought of as a function of $\tau$. The action can then be written as
\ben
 S_0[\phi,\chi] = \frac{1}{2}\sum_J \int {\rm d}\tau\left(\chi'(\tau)\mathcal{K}^{(0)}_J\phi_J^2(\tau)-\mathcal{K}^{(2)}_J\frac{\phi'_J(\tau)^2}{\chi'(\tau)}\right)\,.
\een
where $'$ denotes derivatives with respect to $\tau$. Compared to the previous formulation we have an additional dynamical variable, the ``matter field'' $\chi(\tau)$, and the action has a gauge symmetry under reparametrisations $\tau\rightarrow\tilde\tau(\tau)$ where $\tilde\tau(\tau)$ is any monotonic function of $\tau$. Hence, the new field is a pure gauge degree of freedom; passing to the Hamiltonian picture one now finds a constraint
\ben
\label{repconst}
p_\chi+H^{\rm tot}_\phi := p_\chi -\frac{1}{2} \sum_J\left(\frac{\pi_J^2}{\mathcal{K}_J^{(2)}}+\mathcal{K}^{(0)}_J\phi_J^2\right)\approx 0
\een
where $p_\chi$ is the new canonical momentum conjugate to $\chi$. This constraint generates $\tau$ reparametrisations; in a representation where $\hat{p}_\chi=-{\rm i}\partial_\chi$ acts as a derivative operator, its quantum version is exactly the Schr\"odinger equation defined in the deparametrised approach. The entire construction mimics the usual discussion of parametrised particle systems, here again corresponding to a harmonic oscillator or upside-down harmonic oscillator for each mode $J$. Full details on this construction and everything that follows can be found in \cite{PageWootters}. 

The free parametrised GFT defined in this way is an example of a dynamical system that falls into the ``trinity of relational dynamics'' \cite{trinity}. Its quantum theory can be defined in terms of a Dirac or constraint quantisation where one introduces a kinematical Hilbert space $\mathcal{H}^{{\rm kin}}=\mathcal{H}_\chi\otimes\mathcal{H}_\phi$ subject to (\ref{repconst}). Physical states must solve the constraint, i.e., solve the Schr\"odinger equation of the deparametrised approach. Importantly, $\chi$ is now no longer a classical parameter but corresponds to an operator on $\mathcal{H}^{{\rm kin}}$, and one can work with clock eigenstates or superpositions in $\mathcal{H}_\chi$. This is the reinterpretation of Schr\"odinger time advocated by Page and Wootters \cite{Page-Wootters}. $\chi$ alone is not an observable, since it does not commute with the constraint. Carefully constructing relational observables that do commute with the constraint, one finds a large class of observables of the form $\hat{O}(\chi)$ where $\hat{O}$ is associated to the ``GFT Fock space'' $\mathcal{H}_\phi$, and in particular can be any observable defined in the deparametrised approach. Most importantly, the Dirac quantisation is completely equivalent to the deparametrised Schr\"odinger picture, as the physical Hilbert spaces are identical and all relational observables agree. This is perhaps what would one expect, given that the reparametrisation symmetry has been added by hand into a system that was already deparametrised; but it gives a new conceptual angle on the deparametrised approach and alleviates possible concerns that it might share drawbacks of {\em tempus ante quantum} approaches.

The kinematical Hilbert space $\mathcal{H}^{{\rm kin}}$ constructed here is not the one discussed previously in the algebraic approach, as should be clear from the fact that we now have a separate clock Hilbert space $\mathcal{H}_\chi$; states in $\mathcal{H}_\phi$ can again be interpreted as spin network states living in the Fock space defined in the deparametrised approach. Any physical state is of the form
\ben
|\Psi_{{\rm phys}}\rangle = \int {\rm d}\chi_0\;|\chi_0\rangle\otimes|\psi(\chi_0)\rangle
\een
where $|\psi(\chi_0)\rangle$ solves the Schr\"odinger equation of the deparametrised theory, as is standard in the Page--Wootters picture \cite{trinity}: it encodes an entire history including all possible ``clock readings''.

\section{Conclusion}

The structure of GFT is rather different from other field theories we are used to in theoretical physics: a GFT is a field theory ``not {\em on} but {\em of} spacetime''. In a functional integral, it seems that we are free to sum over arbitrary degrees of freedom, define an action, and try and make the resulting expression well-defined. This was the starting point for quantum descriptions of GFT. However, even this construction is conceptually far from straightforward: first, as we mentioned in the introduction, the link to a classical theory and to other descriptions of quantum mechanics is clearest in the phase space, not in the configuration space path integral. In that sense, a pure configuration space path integral could be seen as a particular proposal for a quantum theory which differs from a phase space definition.\footnote{\label{finalfootnote}For instance, consider a free GFT with a pure mass term, $S_0[\phi]=\frac{1}{2}\int {\rm d}^d g\;\phi^2$. The classical theory has only one solution $\phi=0$, yet the configuration space path integral gives a non-trivial two-point function. The phase space integral instead would reduce the physical phase space to a point $\phi=\pi=0$. This theory is studied in \cite{NewBrunswick} using the configuration space path integral. For higher-derivative theories it is known that such a path integral definition and canonical quantisation may be inequivalent \cite{ghost} and it seems like the same applies to theories without derivatives.} Another point is the definition of a functional integral using either the real exponential $e^{-S}$ or complex amplitude $e^{{\rm i}S}$. In quantum mechanics we are looking for the second type of definition, yet the first one is usually proposed in GFT, so that this is more like a statistical field theory. Since there is no notion of Wick rotation in this context, the relation to a quantum-mechanical definition $e^{{\rm i}S}$ is not particularly clear, which raises the question of whether we should think of GFT as a conventional quantum theory or more a stochastic theory (see \cite{Oppenheim} for a general discussion of path integrals for classical  stochastic and quantum theories, and systems with classical-quantum interactions). Of course, these general conceptual questions are shared with other approaches to quantum gravity \cite{Snowmass}, in particular discrete, background-independent settings. One could also raise more general questions about the meaning of quantum theory when applied to spacetime itself, such as what is the operational meaning of a quantum state or whether there is a role for unitarity and a probability interpretation. Many discussions on these basic questions can be found in the literature, but they are not our main focus here.

In this review we have attempted to summarise various approaches for obtaining a Hilbert space structure for GFT. Such an endeavour seems to implicitly assume that GFT is a conventional quantum theory, and when a link with functional integral techniques is needed we would assume that this is defined with the $e^{{\rm i}S}$ prescription. We then discussed two main approaches: first an {\em algebraic} approach which starts from observables used in the functional integral picture, and tries to encode these as Fock states in a Hilbert space. As we saw, this can be done using additional assumptions, in particular a complex group field and specific commutation relations, but the definition of a physical Hilbert space and its (relational) observables has not been clearly spelled out (though see \cite{NewEdLuca}). We discussed two approaches towards putting these constructions on more solid foundations, which in a sense came from opposing directions: one idea is to take the functional integral as primary and construct a Fock space so that correlation functions agree with the functional integral, another is to reformulate the GFT action as a constrained system so that the Fock space of the algebraic approach emerges as a kinematical Hilbert space in a canonical quantisation. The latter approach is based on techniques also used in loop quantum gravity to define physical states and observables, but the link to spin foam amplitudes is no longer clear since the action needed to be modified. Understanding the precise link between canonical and covariant formulations is a difficult problem also in loop quantum gravity \cite{Zipfel}; perhaps one simply needs to take one of the two as a definition of a theory, as we do in other cases where equivalence between different approaches cannot be proven. 

A second approach uses techniques similar to {\em deparametrisation}; it is available for models with matter coupling to a scalar field $\chi$ which can be used as a clock, whereas the algebraic approach is {\em a priori} independent of a clock choice. Just as for deparametrised models in quantum cosmology and loop quantum gravity, this approach is more straightforward to implement than working with kinematical Hilbert spaces and understanding the possible role of constraints, since there is a conventional notion of evolution in a physical Fock space (and the evolution will be unitary with respect to the chosen clock). In the context of GFT this approach makes specific assumptions about the quadratic part of the action without which it cannot be implemented, but these assumptions are consistent with requirements of renormalisability of the field theory. Conceptually, the usual worries about treating time as classical ({\em tempus ante quantum}) can be addressed by embedding the theory into a larger kinematical structure in which there is a clock Hilbert space associated to $\chi$ and a reparametrisation symmetry (one can even ``de-synchronise'' and introduce a separate clock for each mode \cite{PageWootters}). This means interpreting the Schr\"odinger formulation for GFT in the context of the Page--Wootters formalism. The resulting theories are, at the level of the free theory, equivalent to a standard functional integral definition where it exists. Practically, in either approach it is difficult to add interactions beyond a perturbative regime, and only few studies exist: usually, agreement with late-time cosmology \cite{GFTcosmorel} requires an unstable free theory, and perturbative methods quickly break down. The only exception seems to be interactions that stabilise the group field, and lead to a static cosmology \cite{Condensate}. Again the example of other field theories suggests that numerical simulations and perhaps lattice definitions of GFT should be studied for further progress.
 
\

{\bf Acknowledgements.} --- This work was funded by the Royal Society through the University Research Fellowship Renewal URF$\backslash$R$\backslash$221005. For the purpose of Open Access, the author has applied a CC BY public copyright licence to any Author Accepted Manuscript version arising from this submission. I am grateful to Andrea Calcinari, Luca Marchetti and Daniele Oriti for collaboration and many discussions on many of the issues discussed in this text.

\end{document}